\documentclass[preprint,aps,prd,showpacs,nofootinbib,tightenlines]{revtex4}
\usepackage{amsmath}
\usepackage{amssymb}
\usepackage{epsfig}
\usepackage{graphicx}

\begin{document}

\newcommand{\beq}{\begin{eqnarray}}
\newcommand{\eeq}{\end{eqnarray}}

\newcommand{\non}{\nonumber\\ }
\newcommand{\ov}{\overline}
\newcommand{\rmt}{ {\rm T}}
\newcommand{\psl}{ p \hspace{-2.0truemm}/ }
\newcommand{\qsl}{ q \hspace{-2.0truemm}/ }
\newcommand{\epsl}{ \epsilon \hspace{-2.0truemm}/ }
\newcommand{\nsl}{ n \hspace{-2.2truemm}/ }
\newcommand{\vsl}{ v \hspace{-2.2truemm}/ }

\def \ctp{ {Commun. Theor. Phys. }}
\def \epjc{ {Eur. Phys. J. C }}
\def \jhep{ {J. High Energy Phys. }}
\def \jpg{ {J. Phys. G }}
\def \npb{ {Nucl. Phys. {\bf B}}}
\def \plb{ {Phys. Lett. B }}
\def \prd{ {Phys. Rev. D }}
\def \prl{ {Phys. Rev. Lett. }}
\def \ptp{ {Prog. Theor. Phys. }}
\def \zpc{ {Z. Phys. C }}
\def \cpc{ {Chin. Phys. C }}


\title{Semileptonic decays $B_c^+\to D^{(*)}_{(s)}(l^+\nu_l,l^+l^-,\nu\bar\nu)$
       in the perturbative QCD approach}
\author{Wen-Fei Wang$^{1}$\footnote{wangwf@ihep.ac.cn}, Xin Yu$^{1}$, Cai-Dian L\"{u}$^{1}$,
        and Zhen-Jun Xiao$^{2}$}
\affiliation{$^1$ Center for Future High Energy Physics, Institute of High Energy Physics,\\
Chinese Academy of Sciences, Beijing 100049, People's Republic of China,\\
$^2$ Department of Physics and Institute of Theoretical Physics,\\
Nanjing Normal University, Nanjing, Jiangsu 210023, People's Republic of China}
\date{\today}
\begin{abstract}
In this paper we study the semileptonic decays of $B_c^+\to D^{(*)}_{(s)}(l^+\nu_l,l^+l^-,\nu\bar\nu)$
(here $l$ stands for $e$, $\mu$, or $\tau$). After  evaluating the $B_c^+ \to (D_{(s)},D^*_{(s)})$
transition form factors $F_{0,+,T}(q^2)$ and $V(q^2), A_{0,1,2}(q^2), T_{1,2,3}(q^2)$
by employing the perturbative QCD  factorization approach, we calculate the branching ratios for
all these semileptonic decays. Our predictions for the values of the $B_c^+ \to D_{(s)}$ and
$B_c^+ \to D^*_{(s)}$ transition form factors are consistent with those obtained by using other
methods. The branching ratios of the decay modes with $\bar\nu\nu$ are almost an order of magnitude
larger than the corresponding decays with $l^+l^-$ after the summation over the three neutrino
generations. The branching ratios for the decays with $b\to d$ transitions are much smaller than
those decays with the $b\to s$ transitions, due to the Cabibbo-Kobayashi-Maskawa suppression.
We define ratios $R_D$ and $R_{D^*}$ for the branching ratios with  the $\tau$ lepton versus
$\mu$, $e$ lepton final states to cancel the uncertainties of the form factors, which could possibly be  tested  in the near future.
\end{abstract}

\pacs{13.20.He, 12.38.Bx, 14.40.Nd}

\maketitle

\section{Introduction}\label{sec-intr}

The $B_c$ meson is a pseudoscalar ground state of $b$ and $c$ quarks, and thus the electromagnetic
interaction cannot transform the $B_c$ meson into other hadrons containing $b$ and $c$
quarks. The two difference of quark flavors forbid its annihilation into gluons and being
below the $B-D$ threshold makes the $B_c$ meson stable for strong interaction.
The $B_c$ meson can only decay through weak interactions, so it is an ideal system to
study weak decays of heavy quarks. Either the  heavy quark ($b$ or $c$)
can decay individually,  which makes it different from the $B_{u,d}$ or $B_s$ meson.
The phase space in the $c\to s$ transition is smaller than that in the $b\to c$ transition,
but the Cabibbo-Kobayashi-Maskawa (CKM) matrix element $|V_{cs}|\sim 1$ is much larger
than the CKM matrix element $|V_{cb}|\sim 0.04$. Thus the $c$-quark decays
provide the dominant contribution (about $70\%$) to the decay width of the $B_c$
meson \cite{PAN67-1559}. Because the mass of a $B_c\bar B_c$ pair exceeds
the threshold of $\Upsilon(4S)$, the $B_c$ meson cannot be produced at the $B$ factories.
So comparing with $B_{u,d}$ or $B_s$ meson, the $B_c$ meson decays received much less
experimental attention in the past decades. However, at LHC experiments, around
$5\times10^{10}$ $B_c$ events per year are expected \cite{pap-LHC events,PAN67-1559}
due to the relatively large production cross section, which provides a very good platform
to study various $B_c$ meson decay modes.

Because there is only one hadronic final product, the $B_c$ meson semileptonic decays
among the abundant decay modes are relatively clean in the theoretical treatment.
These semileptonic decays provide good opportunities to measure not only the CKM matrix
elements, such as $|V_{cb}|, |V_{ub}|$, and $|V_{cd}|$, but also the form factors of the
$B_c$ to bottom and charmed mesons transitions. The rare semileptonic decays governed by
the flavor-changing neutral currents are forbidden at tree level in the standard model (SM).
Those decays, which are very sensitive to the contributions of new intermediate particles or
interactions are especially interesting. There are various approaches working on the semileptonic
$B_c$ decays. In Ref.~\cite{arxiv0903.2234}, for example, Dhir and Verma presented a
detailed analysis of the exclusive semileptonic $B_c$ decays in the Bauer-Stech-Wirbel
framework. The authors of the Refs.~\cite{epjcd4-18,prd65-094037,arxiv1006-4231} studied the
semileptonic $B_c$ decays in the relativistic and/or constituent quark model. In Refs.
~\cite{M-Jamil-Aslam,M-J-Aslam}, $B_c\to D^*_sl^+l^-$ decays were studied in the SM with the
fourth-generation and supersymmetric models. The three point QCD sum rules approach was adopted
to investigate the $B_c^+\to D^{*+}_{(s)} l^+l^-$ in \cite{prd77-114024} and
$B_c^+\to D_{(s)}^+(l^+l^-,\bar\nu\nu)$ in \cite{prd78-036005}.

In this paper, we will study the semileptonic decays of
$B_c^+\to D^{(*)}_{(s)}(l^+\nu_l,l^+l^-,\nu\bar\nu)$ (here $l$ stands for leptons $e, \mu,$ or $\tau$)
the perturbative QCD (pQCD) approach \cite{pap-pQCD}.  These semileptonic decays are governed by the
form factors. At the maximum recoil region, the final state meson is collinear with a large momentum.
he spectator $c$ quark in $B_c$ meson thus needs a hard gluon to kick it off from almost zero
momentum to a collinear state. However, when doing integrations of momentum fractions of valence
quarks, endpoint singularity occurs. A natural way to kill this singularity is to pick up the
neglected transverse momentum in the collinear factorization. With the additional transverse momentum
cale $k_T$ \cite{pqcd}, double logarithms appear in the calculation. We have to use the renormalization
group equation to perform the resummation resulting in the so-called Sudakov form factors
\cite{Suda-factor} and make the perturbative calculation of the hard amplitudes (form factors)
infrared safe. The pQCD approach is widely adopted to calculate the transition form factors
of $B_{u,d}$ and $B_s$ meson\cite{prd86-114025,prd87-097501,pQCD-FF}.
Furthermore, various $B_c$ decay modes have also been studied in Refs.
\cite{cpc37-093102,PRD81-014022-EPJC45-711-EPJC60-107} in the pQCD approach.

The structure of this paper is as follows. After this Introduction, we collect the
distribution amplitudes of the $B_c$, $D^{(*)}$ and $D_s^{(*)}$ mesons in Sec. II.
Based on the $k_{\rm T}$ factorization formalism, we calculate and present
the expressions for the $B_c \to (D^{(*)}, D_s^{(*)})$ transition form factors in
the large recoil regions in Section III.
The numerical results and relevant discussions are given in Sec. IV. And Sec. V
contains  a short summary.

\section{Kinematics and the wave functions}\label{sec-wfun}

\begin{figure}[tbp]
\vspace{-1cm}
\centerline{\epsfxsize=9cm \epsffile{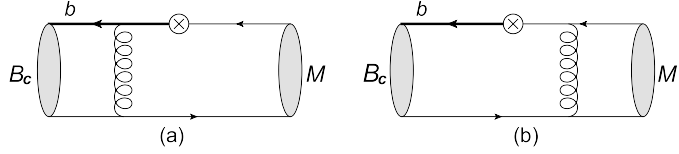}}
\caption{The leading-order Feynman diagrams for the transition of $B_c^+\to(D^{(*)}, D_s^{(*)})$,
  where $M$ stands for a $D^{(*)}$ or $D_s^{(*)}$ meson, and $\otimes$ is the weak vertex.}
\label{fig-fig1}
\end{figure}

The lowest-order diagrams for $B_c^+\to(D^{(*)}, D_s^{(*)})$ transitions are displayed in
Fig.~\ref{fig-fig1}, where $M$ stands for $D^{(*)}$ or $D_s^{(*)}$
meson, the $\otimes$ is the weak vertex for the leptonic pairs to come out. In the rest frame of $B_c$ meson, with the
$m_{B_c}$ standing for the mass of the $B_c$ meson, and $m$ for the $D_{(s)}$ or
$D^*_{(s)}$ mesons, the momenta of $B_c$ and $D^{(*)}_{(s)}$ mesons are defined in the
light-cone coordinates as~\cite{prd67-054028,cpc37-093102}
\beq\label{eq-mom-p1p2}
p_1=\frac{m_{B_c}}{\sqrt{2}}(1,1,0_\bot),\quad
p_2=\frac{m_{B_c}}{\sqrt{2}}(r\eta^+,r\eta^-,0_\bot),
\eeq
with $r=m/m_{B_c}$ and $\eta^\pm=\eta\pm\sqrt{\eta^2-1}$.
As for the $\eta$ in $\eta^\pm$, the expression
\beq
\eta=\frac{1}{2r}\left[1+r^2-\frac{q^2}{m^2_{B_c}}\right]
\eeq
can be evaluated from $q^2= (p_1-p_2)^2$ which is the invariant mass of the lepton pairs.
The momenta of the spectator quarks in the $B_c$ and $D^{(*)}_{(s)}$ mesons are parameterized as
\beq\label{eq-k1k2}
k_1 =(x_1\frac{m_{B_c}}{\sqrt{2}},x_1\frac{m_{B_c}}{\sqrt{2}},k_{1\bot}), \quad
k_2=(x_2\frac{m_{B_c}}{\sqrt{2}}r\eta^+,
x_2\frac{m_{B_c}}{\sqrt{2}}r\eta^-,k_{2\bot}).
\eeq
For the $D^*_{(s)}$ mesons, we define their polarization vector
$\epsilon$ as
\beq\label{eq-def-epsilon}
\epsilon_L=\frac{1}{\sqrt2}(\eta^+,-\eta^-,0_\bot), \quad
\epsilon_T=(0,0,1),
\eeq
where $\epsilon_L$ and $\epsilon_T$ denote the longitudinal and
transverse  polarization of the $D^*_{(s)}$ mesons, respectively.

In this work, we use the same distribution amplitude for the $B_c$ meson
as that used in Refs.~\cite{cpc37-093102,epjc45-711,prd81-014002,prd87-074027},
\beq\label{eq-Bc-wavefun}
\Phi_{B_c}(x)=\frac{i}{\sqrt{2N_c}}[(\psl+m_{B_c})\gamma_{5}
\phi_{B_c}(x)]_{\alpha\beta}\;,
\eeq
with
\beq\label{eq-Bc-wave}
\phi_{B_c}(x)= \frac{f_{B_c}}{2\sqrt{2N_c}}\delta(x-m_c/m_{B_c})
\exp[-\omega^2_{B_c}b^2/2]\;,
\eeq
where $m_c$ is the mass of $c$-quark.
Because the $B_c$ meson consists of two heavy quarks $b$ and $c$, just like a heavy quarkonium, the
non-relativistic QCD framework can be applied, which means the leading-order wave function should
be just the zero-point wave function  shown in Eq.~(\ref{eq-Bc-wave}).

For the $D^{(*)}_{(s)}$ mesons, up to twist-3 accuracy, the two-parton light-cone distribution
amplitudes are defined as \cite{prd67-054028,prd78-014018}
\beq
\langle D_{(s)}(p)|q_\alpha(z)\bar c_\beta(0)|0\rangle&=&
\frac{i}{\sqrt{2N_C}}
\int_0^1dx e^{ixp\cdot z}\left[\gamma_5\left(\psl+m\right)
\phi_{D_{(s)}}(x,b)\right]_{\alpha\beta}\;,\non
\langle D_{(s)}^*(p)|q_\alpha(z)\bar c_\beta(0)|0\rangle&=&
-\frac{1}{\sqrt{2N_C}}\int_0^1dx e^{ixp\cdot z}
\big[\epsl_L(\psl+m)\phi_{D^*_{(s)}}(x,b)\non
&& +\;\epsl_T(\psl+m)\phi_{D^*_{(s)}}(x,b)\big]_{\alpha\beta}\;,
\eeq
with
\beq
  \int_0^1 dx\phi_{D_{(s)}}(x,0)&=&\frac{f_{D_{(s)}}}{2\sqrt{2N_c}}\;,\non
  \int_0^1dx\phi_{D^*_{(s)}}(x,0)&=&\frac{f_{D^*_{(s)}}}{2\sqrt{2N_c}}\;,
\eeq
as the normalization conditions.
We adopt $f_D=206.7\pm8.9$ MeV and $f_{D_s}=260.0\pm5.6$ MeV in PDG \cite{pdg2012}
by experimental average for $D$ and $D_s$ mesons, respectively. For  the $D^*$ or $D^*_s$ meson,
we adopt the same decay constant and distribution amplitude for the longitudinal and transverse components.
Since there is no experimental data, we use $f_{D^*}=270$ MeV and $f_{D^*_s}=310$ MeV for $D^*$ and
$D^*_s$ meson considering of the results in Refs.~\cite{decay-cons} and assume a $10\%$ uncertainty.
The distribution amplitude for the $D_{(s)}$ meson is
\beq
\phi_{D_{(s)}}=\frac{1}{2\sqrt{2N_C}}f_{D_{(s)}}6x(1-x)
\left[1+C_{D_{(s)}}(1-2x)\right]\exp\left[-\frac{\omega_{D_{(s)}}^2b^2}{2}\right],
\eeq
which is a $k_T$-dependent form with $C_D=0.5, \omega_D=0.1$ and $C_{D_s}=0.4, \omega_{D_s}=0.2$
for $D$ and $D_s$ mesons, respectively \cite{prd78-014018}. In this work, we also adopt the same
distribution amplitude for both the vector meson  $D^*_{(s)}$ and pseudoscalar meson $D_{(s)}$ because
of their small mass difference \cite{prd78-014018}.

\section{Form factors of semileptonic decays}\label{sec-ff}

The form factors $F_+(q^2), F_0(q^2)$ for the $B_c$ to pseudoscalar meson $D_{(s)}$ transition
induced by the vector current can be defined as \cite{zpc29-637,npb592-3}
\beq\label{eq-def-F+F0}
\langle D_{(s)}(p_2)|\bar{q}(0)\gamma_{\mu}b(0)|B_c(p_1)\rangle&=&
\left[(p_1+p_2)_{\mu}-\frac{m_{B_c}^2-m^2}{q^2}q_{\mu}\right]
F_+(q^2)\non
&+&\frac{m_{B_c}^2-m^2}{q^2}q_{\mu}F_0(q^2),
\eeq
where $q=p_1-p_2$ is the momentum of the lepton pairs. In order to cancel the poles at
$q^2=0$, $F_+(0)$ should be equal to $F_0(0)$. For the sake of convenience, we define the auxiliary
form factors $f_1(q^2)$ and $f_2(q^2)$,
\beq\label{eq-def-f1f2}
\langle D_{(s)}(p_2)|\bar{q}(0)\gamma_{\mu}b(0)|B_c(p_1)\rangle=
f_1(q^2)p_{1\mu}+f_2(q^2)p_{2\mu}.
\eeq
In terms of $f_1(q^2)$ and $f_2(q^2)$ the form factors $F_+(q^2)$ and $F_0(q^2)$ are
\beq\label{eq-f+f0}
F_+(q^2)&=&\frac12\left[f_1(q^2)+f_2(q^2)\right], \notag\\
F_0(q^2)&=&\frac12 f_1(q^2)\left[1+\frac{q^2}{m_{B_c}^2-m^2}\right]
+\frac12 f_2(q^2)\left[1-\frac{q^2}{m_{B_c}^2-m^2}\right].
\eeq
The form factor $F_T(q^2)$ for the $B_c\to D_{(s)}$ transition
induced by the tensor current can be defined as \cite{npb592-3}
\beq\label{eq-def-FT}
\langle D_{(s)}(p_2)|\bar{q}(0)\sigma_{\mu\nu} b(0)|B_c(p_1)\rangle=
i\left[p_{2\mu}q_\nu-q_\mu p_{2\nu}\right]\frac{2F_T(q^2)}{m_{B_c}+m}
\;.
\eeq

There are seven form factors $V(q^2),A_{0,1,2}(q^2)$ and $T_{1,2,3}(q^2)$
needed for the transition of $B_c\to D^*_{(s)}$ in this work.
The form factors $V(q^2)$ and $A_{0,1,2}(q^2)$ are defined
by~\cite{npb592-3,epjc28-515,prd65-014007}
\beq\label{eq-def-VA012}
\langle D^*_{(s)}(p_2)|\bar q(0)\gamma_\mu b(0) |B_c (p_1)\rangle
&=&\epsilon_{\mu\nu\alpha\beta}\epsilon^{\nu*} p_1^\alpha p_2^\beta
\frac{2V(q^2)}{m_{B_c}+m}\;,\\
\langle D^*_{(s)}(p_2)|\bar q(0) \gamma_\mu  \gamma_5 b(0)|B_c(p_1)  \rangle
&=&i\left[ \epsilon_\mu^* -\frac{\epsilon^*\cdot q}{q^2}q_\mu\right]
(m_{B_c}+ m) A_1(q^2) \non
&-&i\left[ (p_1+p_2)_\mu -\frac{m_{B_c}^2 -m^2}{q^2}q_\mu\right]
(\epsilon^* \cdot q) \frac {A_2(q^2) }{m_{B_c}+ m}\non
&+&i\frac{2m(\epsilon^* \cdot q)}{q^2} q_\mu A_0(q^2) ,
\eeq
where $\epsilon^*$  is the polarization vector of the $D^*_{(s)}$ meson.
The form factors $T_{1,2,3}$ are defined by
\cite{npb592-3,prd53-3672}
\beq
\langle D^*_{(s)}(p_2)|\bar q(0)\sigma_{\mu\nu}q^\nu(1&+&\gamma_5)
b(0)|B_c(p_1)  \rangle= i\epsilon_{\mu\nu\alpha\beta}\epsilon^{*\nu}
p_1^\alpha p_2^\beta 2 T_1(q^2)\non
&+&\left[\epsilon^*_\mu(m^2_{B_c}-m^2)-(\epsilon^*\cdot q)(p_1+p_2)_\mu
\right]T_2(q^2)\non
&+&(\epsilon^*\cdot q)\left[q_\mu-\frac{q^2}{m^2_{B_c}-m^2}(p_1+p_2)_\mu
\right]T_3(q^2)\;,
\eeq
with $T_1(0)=T_2(0)$ implied by the identity
\beq\label{eq-condition-t1t2}
\sigma_{\mu\nu}\gamma_5=-\frac{i}{2}\epsilon_{\mu\nu\alpha\beta}
\sigma^{\alpha\beta}\;.
\eeq

In the transverse configuration $b$-space and by
including the Sudakov form factors and the threshold resummation
effects, we obtain the $B_c\to D_{(s)}$ form factors
$f_1(q^2), f_2(q^2)$ and $F_T(q^2)$ as follows
\beq  
f_1(q^2)&=&16\pi m_{B_c}^2rC_F\int dx_1 dx_2\int b_1 db_1 b_2 db_2
\;\phi_{B_c}(x_1) \phi_{D_{(s)}}(x_2,b_2) \non
&\times &\Bigl\{\left[ 1-rx_2\right]\cdot h_1(x_1,x_2,b_1,b_2)
\cdot \alpha_s(t_1)\exp\left [-S_{ab}(t_1) \right ] \non
&-& \left[r+2x_1(1-\eta)\right]\cdot h_2(x_1,x_2,b_1,b_2)
\cdot \alpha_s (t_2)\exp\left [-S_{ab}(t_2) \right] \Bigr \}\;,
\label{eq:f1q2}
\eeq
\beq
f_2(q^2)&=&16\pi m_{B_c}^2C_F\int dx_1 dx_2\int b_1 db_1 b_2 db_2 \;
\phi_{B_c}(x_1) \phi_{D_{(s)}}(x_2,b_2) \non
&\times& \Bigl\{ \left[1-2rx_2(1-\eta)\right]
\cdot h_1(x_1,x_2,b_1,b_2)\cdot \alpha_s(t_1)
 \exp\left [-S_{ab}(t_1) \right ] \non
&+& \left[ 2r-x_1\right] \cdot h_2(x_1,x_2,b_1,b_2)
\cdot \alpha_s (t_2)\exp\left [-S_{ab}(t_2) \right] \Bigr \}\;,
\label{eq:f2q2}
\eeq
\beq
F_T(q^2)&=&8\pi m_{B_c}^2C_F(1+r)\int dx_1 dx_2\int b_1 db_1 b_2 db_2 \;
\phi_{B_c}(x_1) \phi_{D_{(s)}}(x_2,b_2) \non
&\times& \Bigl\{ \left[1-rx_2\right]
\cdot h_1(x_1,x_2,b_1,b_2)\cdot \alpha_s(t_1)
 \exp\left [-S_{ab}(t_1) \right ] \non
&+& \left[ 2r-x_1\right] \cdot h_2(x_1,x_2,b_1,b_2)
\cdot \alpha_s (t_2)\exp\left [-S_{ab}(t_2) \right] \Bigr \}\;,
\label{eq:FTq2}
\eeq
where $C_F=4/3$ is the color factor. The functions $h_1$ and $h_2$, the scales $t_1$, $t_2$ and the
Sudakov factors $S_{ab}$ are the same as those given in Refs.~\cite{prd67-054028,cpc37-093102}.

The expressions of form factors $V(q^2),A_{0,1,2}(q^2)$
and $T_{1,2,3}(q^2)$ for the $B_c\to D^*_{(s)}$ transition in the pQCD approach are:
\beq  
V(q^2)&=&8\pi m_{B_c}^2C_F(1+r)\int dx_1 dx_2\int b_1 db_1 b_2 db_2
 \;\phi_{B_c}(x_1)\phi^T_{D^*_{(s)}}(x_2,b_2)\non
&\times & \Bigl \{\left[1-rx_2\right]
\cdot h_1(x_1,x_2,b_1,b_2)\cdot \alpha_s(t_1)
 \exp\left [-S_{ab}(t_1) \right ] \non
&+& r\cdot h_2(x_1,x_2,b_1,b_2)
\cdot \alpha_s (t_2)\exp\left [-S_{ab}(t_2) \right] \Bigr \}\;,
\label{eq:Vqq}
\eeq
\beq
A_0(q^2)&=&8\pi m_{B_c}^2C_F\int dx_1 dx_2\int b_1 db_1 b_2 db_2
 \;\phi_{B_c}(x_1)\phi^L_{D^*_{(s)}}(x_2,b_2)\non
&\times & \Bigl \{  \left[1-rx_2(r-2\eta)
+r\left(1-2x_2\right)\right]\non
&\times& h_1(x_1,x_2,b_1,b_2)\cdot \alpha_s(t_1)
 \exp\left [-S_{ab}(t_1) \right ] \non
&+& \left [r^2+x_1(1-2r\eta)\right ]\cdot h_2(x_1,x_2,b_1,b_2)
\cdot \alpha_s (t_2)\exp\left [-S_{ab}(t_2) \right] \Bigr \}\;,
\label{eq:A0qq}
\eeq
\beq
A_1(q^2)&=&16\pi m_{B_c}^2C_F\frac{r}{1+r}
\int dx_1 dx_2\int b_1 db_1 b_2 db_2
\;\phi_{B_c}(x_1)\phi^T_{D^*_{(s)}}(x_2,b_2)\non
&\times & \Bigl \{\left [1+rx_2\eta
-2rx_2+\eta\right ]
\cdot h_1(x_1,x_2,b_1,b_2)\cdot \alpha_s(t_1)\exp[-S_{ab}(t_1)]\non
& + &\left[r\eta -x_1 \right]\cdot h_2(x_1,x_2,b_1,b_2)
\cdot \alpha_s (t_2)\exp[-S_{ab}(t_2)] \Bigr \}\;,
\label{eq:A1qq}
\eeq
\beq
A_2(q^2)&=&\frac{(1+r)^2(\eta-r)}{2r(\eta^2-1)}\cdot A_1(q^2)-
8\pi m_{B_c}^2C_F\frac{1+r}{\eta^2-1}\int dx_1 dx_2\int b_1 db_1 b_2 db_2
\;\phi_{B_c}(x_1)\non
&\times & \phi^L_{D^*_{(s)}}(x_2,b_2)\cdot
\Bigl \{  \left[\eta(1-r^2x_2)-rx_2(1-2\eta^2-2r) +(1-r)
-r\eta(1+2x_2)\right]\non
&\times&
h_1(x_1,x_2,b_1,b_2)\cdot \alpha_s(t_1)
 \exp\left [-S_{ab}(t_1) \right ] \non
&+& \left[ r(1-x_1+2x_1\eta^2)-\eta(r^2+x_1)\right]\non
&\times& h_2(x_1,x_2,b_1,b_2)
\cdot \alpha_s (t_2)\exp\left [-S_{ab}(t_2) \right] \Bigr \}\;,
\label{eq:A2qq}
\eeq
\beq
T_1(q^2)&=&8\pi m_{B_c}^2C_F\int dx_1 dx_2\int b_1 db_1 b_2 db_2
 \;\phi_{B_c}(x_1)\phi^T_{D^*_{(s)}}(x_2,b_2)\non
&\times & \Bigl \{  \left[1+r(1-x_2(2+r-2\eta))\right]
\cdot h_1(x_1,x_2,b_1,b_2)\cdot \alpha_s(t_1)
 \exp\left [-S_{ab}(t_1) \right ] \non
&+& r\left[1- x_1\right]\cdot h_2(x_1,x_2,b_1,b_2)
\cdot \alpha_s (t_2)\exp\left [-S_{ab}(t_2) \right] \Bigr \}\;,
\label{eq:T1qq}
\eeq
\beq
T_2(q^2)&=&16\pi m_{B_c}^2C_F\frac{r}{1-r^2}
\int dx_1 dx_2\int b_1 db_1 b_2 db_2
 \;\phi_{B_c}(x_1)\phi^T_{D^*_{(s)}}(x_2,b_2)\non
&\times & \Bigl \{  \left[(1-r)(1+\eta)+2rx_2(r-\eta)
+rx_2(2\eta^2-r\eta-1)  \right]\non
&\times& h_1(x_1,x_2,b_1,b_2)\cdot \alpha_s(t_1)
 \exp\left [-S_{ab}(t_1) \right ] \non
&+& \left[r(1+x_1)\eta -r^2-x_1 \right]\cdot h_2(x_1,x_2,b_1,b_2)
\cdot \alpha_s (t_2)\exp\left [-S_{ab}(t_2) \right] \Bigr \}\;,
\label{eq:T2qq}
\eeq
\beq
T_3(q^2)&=&\frac{r+\eta}{r}\cdot\frac{1-r^2}{2(\eta^2-1)}
\cdot T_2(q^2)-\frac{1-r^2}{(\eta^2-1)}\non
&\times&8\pi m_{B_c}^2C_F
\int dx_1 dx_2\int b_1 db_1 b_2 db_2
 \;\phi_{B_c}(x_1)\phi^L_{D^*_{(s)}}(x_2,b_2)\non
&\times & \Bigl \{  \left[1+rx_2(\eta-2)+\eta \right]
\cdot h_1(x_1,x_2,b_1,b_2)\cdot \alpha_s(t_1)
 \exp\left [-S_{ab}(t_1) \right ] \non
&+& \left[x_1\eta-r \right]\cdot h_2(x_1,x_2,b_1,b_2)
\cdot \alpha_s (t_2)\exp\left [-S_{ab}(t_2) \right] \Bigr \}\;.
\label{eq:T3qq}
\eeq
One should note that the expressions for the form factors
$f_{1,2}(q^2)$, $F_T(q^2)$, $V(q^2)$, $A_{0,1,2}(q^2)$  and
$T_{1,2,3}(q^2)$ given in Eqs. (\ref{eq:f1q2})-(\ref{eq:T3qq})
are the results at leading order of the pQCD approach.
The next-to-leading-order contributions  to the form
factors of $B \to (\pi, K, \eta^{(\prime)})$ transitions
given in Refs. \cite{prd85-074004,prd86-114025,prd87-097501}
are not available here because of the large mass of $c$-quark and
$(D_{(s)}, D^*_{(s)})$ mesons.

One should note that the pQCD predictions for the considered
form factors are reliable only for the small
values of $q^2$. For the form factors in the large-$q^2$ region,
one has to make an extrapolation for them from the low-$q^2$
region to large-$q^2$ region. In this work we make the
extrapolation by using the formula in
Refs.~\cite{cpc37-093102,prd79-054012}
\beq
F(q^2)=F(0)\cdot \exp{\left[a\cdot q^2+b\cdot (q^2)^2\right]},
\eeq
where $F$ stands for the form factors $F_{0,+,T}, V, A_{0,1,2}$ and
$T_{1,2,3}$, and $a, b$ are the constants to be determined by the
fitting procedure.

The $B_c^-\to \bar D^0l^-\bar \nu_l$ and
$B_c^-\to \bar D^{*0}l^-\bar \nu_l$ decays are from the quark level  $b\to ul^-\bar \nu$  charged current
transition. The effective Hamiltonian for such transition is
\cite{eff-ham}
\beq  \label{eq-hamiltonian}
{\cal H}_{eff}(b\to ul^-\bar \nu_l)=\frac{G_F}{\sqrt{2}}V_{ub}\;
\bar{u} \gamma_{\mu}(1-\gamma_5)b \cdot \bar l\gamma^{\mu}
(1-\gamma_5)\nu_l,
\eeq
where $G_F=1.16637\times10^{-5} GeV^{-2}$ is the Fermi-coupling constant and
$V_{ub}$ is one of the CKM matrix elements. With the form factors calculated in
Eqs.~(\ref{eq:f1q2},\ref{eq:f2q2},\ref{eq:Vqq}-\ref{eq:A2qq}),
one can easily get the  differential decay width
expression for $B_c^-\to \bar D^0l^-\bar \nu_l$ and
$B_c^-\to \bar D^{*0}l^-\bar \nu_l$.

For those flavor-changing neutral-current one-loop decay modes, such as
$B_c\to D^{(*)}l^+l^-$ and $B_c\to D_s^{(*)}l^+l^-$, are transitions of
$b\to dl^+l^-$ and $b\to sl^+l^-$ at quark level, respectively.
The effective Hamiltonians and the
corresponding differential decay widths  are more complicated, we refer the readers to
Refs. \cite{prd86-114025,prd53-3672,npb612-25,epjc40-565,prd61-074024}.

For the decay modes of $B_c\to D^{(*)}_s\nu\bar\nu$, the effective
Hamiltonian is \cite{eff-ham}
\beq
\label{eq-Ham-nunu}
{\cal H}_{eff}(b \to s\nu \bar \nu)&=& \frac{G_F}{\sqrt{2}}
\frac{\alpha_{em}}{2 \pi \sin^2(\theta_W)}
V_{tb} V_{ts}^* \eta_X X(x_t) \;
\left[ {\bar s}\gamma^\mu (1-\gamma_5) b \right ]
\left[ {\bar \nu} \gamma_\mu(1-\gamma_5) \nu \right]
\eeq
where $\theta_W$ is the Weinberg angle with $\sin^2(\theta_W)=0.231$
\cite{pdg2012}, $V_{tb}$ and $V_{ts}$ are CKM matrix elements and
$\alpha_{em}\approx 1/137$ is the fine structure constant.
The function $X(x_t)$ can be found in
Ref.~\cite{eff-ham}, while $\eta_X\approx 1$ is the QCD radiative correction factor \cite{eff-ham}.
As for the decay modes of $B_c\to D^{(*)}\nu\bar\nu$, their effective Hamiltonian
can be obtained by a simple replacement of $s\to d$ in Eq. (\ref{eq-Ham-nunu}).
The corresponding differential decay widths for $B_c\to D_{(s)}\nu\bar\nu$
is the same as $B\to \pi(K)\nu\bar\nu$ in Ref.~\cite{prd86-114025} except
the replacements $m_B\to m_{B_c}$ and $m_P\to m$.
While for the decay modes of $B_c\to D^*_s\nu\bar\nu$, the differential
decay width is~\cite{jhep0707-072}
\beq\label{eq-dVdvv}
\frac{d\Gamma(B_c\to D^*_s\nu\bar\nu)}{dq^2}&=&\frac{G^2_F\alpha^2_{em}}
{2^{10}\pi^5m^3_{B_c}}\cdot \left|\frac{X(x_t)}{sin^2(\theta_w)}\right|^2
\cdot\eta^2_X\cdot\left|V_{tb}V^*_{ts}\right|^2\lambda^{\frac12}
\bigg\{8\lambda q^2\frac{V^2}{(m_{B_c}+m)^2}\non
&&+\frac{1}{m^2}\bigg[\lambda^2\frac{A^2_2}{(m_{B_c}+m)^2}+
(m_{B_c}+m)^2(\lambda+12m^2q^2)\cdot A^2_1\non
&&-2\lambda(m^2_{B_c}-m^2-q^2)\cdot Re[A^*_1A_2] \bigg]\bigg\}\;,
\eeq
where $V, A_1$ and $A_2$ are the form factors of $B_c\to D^*_s$ transition,
and the phase-space factor
\beq
\lambda=(m^2_{B_c}+m^2-q^2)^2-4m^2_{B_c}m^2\;.
\eeq

\section{Numerical results and discussions} \label{sec-num}

In the numerical calculations we adopt the following input parameters
\cite{pdg2012}
\beq
m_{B^-_c}&=&6.277~{\rm GeV}, \quad
m_{\bar D^0}=1.865~{\rm GeV}, \quad m_{D^-}=1.870~{\rm GeV},\non
m_{\bar D^{*0}}&=&2.007~{\rm GeV}, \quad m_{D^{*-}}=2.010~{\rm GeV},
\quad m_{D_s^{-}}=1.969~{\rm GeV},\non
m_{\bar D_s^{*-}}&=&2.112~{\rm GeV}, \quad m_{\tau}=1.777~{\rm GeV},
\quad m_{c}=1.275\pm0.025~{\rm GeV},\non
\tau_{B_c}&=&(0.45\pm0.04)~{\rm ps},
\label{eq-inputs}
\eeq
For the CKM matrix element $V_{ub}$, we adopt the value in
Refs.~\cite{prd86-114025,babar-Vub}. And we use
$|V_{tb}|=0.999, |V_{ts}|=0.0404$ and $|V_{td}/V_{ts}|=0.211$ \cite{pdg2012} in this work.
As for the decay constant of the $B_c$ meson, we adopt $0.489~{\rm GeV}$ \cite{plb651-171}
as its central value, and give it an uncertainty of $0.050~{\rm GeV}$.

The numerical values of the $B_c\to D$ and $B_c\to D_s$ transition
form factors $F_{0,+,T}$ at $q^2=0$ and their fitted parameters $a, b$ are
listed in Table \ref{tab-FF-I}.
The numerical values of the form factors $V, A_{0,1,2}$ and
$T_{1,2,3}$ at $q^2=0$ for the $B_c\to D^*$ and $B_c\to D^*_s$
transitions are listed in Table \ref{tab-FF-II}.
The first error of the pQCD predictions for the form factors in Table
\ref{tab-FF-I} and Table \ref{tab-FF-II} is induced by the $B_c$ meson wave function parameter
$\omega_{B_c}=1.0\pm0.1$;
the second error comes from the uncertainty of decay constant $f_{B_c}$;
the third error comes from the uncertainty of decay constants of the $D^{(*)}_{(s)}$ mesons;
the fourth error in Tables \ref{tab-FF-I} and \ref{tab-FF-II}
comes from the uncertainty of $D_{(s)}^{(*)}$ wave function $C_{D^{(*)}}=0.5\pm0.1$ or $C_{D^{(*)}_s}=0.4\pm0.1$;
the fifth error comes from $m_c=1.275\pm0.025$ GeV.
The errors from the uncertainty of $\omega_{D^{(*)}}=0.10\pm0.02$ or
$\omega_{D^{(*)}_s}=0.20\pm0.04$ are very small that have been neglected.

Unlike the form factors at maximum  recoil, the extrapolation parameters $a$, $b$ of the form
factors are less sensitive to the decay constant and wave function of $D_{(s)}^{(*)}$ meson.
In Tables \ref{tab-FF-I} and \ref{tab-FF-II}, we only show uncertainties for  the parameter $a$
and  $b$  from $B_c$ meson wave function parameter $\omega_{B_c}$,
and  from quark mass uncertainty $m_c=1.275\pm0.025$ GeV.
As a comparison, we also present some results obtained by other authors
based on different methods in Table \ref{tab-FF-literature}.
It is easy to see that our results
are consistent  with the results in literature.

\begin{table}[thb]
\begin{center}
\caption{The pQCD predictions for form factors $F_0, F_+$ and $F_T$
at $q^2=0$ and the parametrization constants $a$ and $b$
for $B_c\to D$  and $B_c\to D_s$ transitions.}
\label{tab-FF-I}
\begin{tabular}{c c c c} \hline\hline
\  \ &$ \ \ F(0)$ &$\ \ a$ &$ \ \ b$  \\
 \hline
\
$F_0^{B_c\to D}$ \  &\ $0.19\pm0.02\pm0.02\pm0.01\pm0.01\pm0.01$&\ \ $0.038\pm0.001\pm0.000$
& \ \  $0.0013\pm0.0001$ \; \\
\
$F_+^{B_c\to D}$ \ &\ $0.19\pm0.02\pm0.02\pm0.01\pm0.01\pm0.01$ &\ \ $0.059\pm0.001\pm0.001$
&\ \ $0.0020\pm0.0001$ \; \\
\
$F_T^{B_c\to D}$ \ &\ $0.20\pm0.02\pm0.02\pm0.01\pm0.01\pm0.01$ &\ \ $0.070\pm0.001\pm0.001$
 &\ \ $0.0021^{+0.0000}_{-0.0001}$ \; \\ \hline
\
$F_0^{B_c\to D_s}$ \ &\ $0.27\pm0.03\pm0.03\pm0.02\pm0.01\pm0.01$
&\ \ $0.039\pm0.002\pm0.001$ &\ \ $0.0015^{+0.0001}_{-0.0000}$ \; \\
\
$F_+^{B_c\to D_s}$ \ &\ $0.27\pm0.03\pm0.03\pm0.02\pm0.01\pm0.01$
&\ \ $0.061\pm0.002\pm0.001$&\ \ $0.0023^{+0.0001}_{-0.0000}$ \; \\
\
$F_T^{B_c\to D_s}$ \ &\ $0.28\pm0.03\pm0.03\pm0.02\pm0.01\pm0.01$
&\ \ $0.073\pm0.002\pm0.001$&\ \ $0.0025^{+0.0000}_{-0.0001}$ \;  \\ \hline \hline
\end{tabular}
\end{center}
\end{table}

\begin{table}[thb]
\begin{center}
\caption{The pQCD predictions for form factors $ A_{0,1,2},V$ and
$T_{1,2,3}$ at $q^2=0$ and the parametrization constants $a$ and $b$
for $B_c\to D^*$  and $B_c\to D^*_s$ transitions.}
\label{tab-FF-II}
\begin{tabular}{c c c c} \hline\hline
\ &$ \  F(0)$ &$\  a$ &$ \  b$  \\ \hline
$A_0^{B_c\to D^*}$ &\ $0.17\pm0.02\pm0.02\pm0.02\pm0.01\pm0.00$ &\ $0.063\pm0.001\pm0.001$
&\ $0.0024\pm0.0000\pm0.0000$  \\
$A_1^{B_c\to D^*}$  &\ $0.18\pm0.02\pm0.02\pm0.02\pm0.01\pm0.01$ &\  $0.043\pm0.001\pm0.001$
&\  $0.0018\pm0.0001\pm0.0001$  \\
$A_2^{B_c\to D^*}$  &\ $0.20\pm0.02\pm0.02\pm0.02\pm0.01\pm0.01$ &\  $0.067\pm0.001\pm0.001$
&\  $0.0026\pm0.0001\pm0.0001$  \\
$V^{B_c\to D^*}$  &\ $0.25\pm0.03\pm0.03\pm0.03\pm0.01\pm0.01$&\  $0.073\pm0.002\pm0.001$
& \   $0.0029\pm0.0001\pm0.0001$ \\
$T_1^{B_c\to D^*}$  &\ $0.22\pm0.02\pm0.02\pm0.02\pm0.01\pm0.01$ &\  $0.063\pm0.001\pm0.001$
&\  $0.0027\pm0.0001\pm0.0001$  \\
$T_2^{B_c\to D^*}$  &\ $0.22\pm0.02\pm0.02\pm0.02\pm0.01\pm0.01$ &\  $0.038\pm0.001\pm0.001$
&\  $0.0017\pm0.0001\pm0.0001$  \\
$T_3^{B_c\to D^*}$  &\ $0.20\pm0.02\pm0.02\pm0.02\pm0.01\pm0.01$ &\  $0.077\pm0.002\pm0.001$
&\  $0.0049\pm0.0001\pm0.0001$  \\
\hline
$A_0^{B_c\to D^*_s}$  &\ $0.21\pm0.02\pm0.02\pm0.02\pm0.01\pm0.01$ &\  $0.064\pm0.001\pm0.001$
&\  $0.0031\pm0.0002\pm0.0001$  \\
$A_1^{B_c\to D^*_s}$  &\ $0.23\pm0.02\pm0.02\pm0.02\pm0.01\pm0.01$ &\  $0.044\pm0.002\pm0.001$
&\  $0.0022\pm0.0002\pm0.0001$  \\
$A_2^{B_c\to D^*_s}$  &\ $0.25\pm0.03\pm0.03\pm0.03\pm0.01\pm0.01$ &\  $0.069\pm0.002\pm0.001$
&\  $0.0035\pm0.0002^{+0.0002}_{-0.0001}$ \\
$V^{B_c\to D^*_s}$  &\ $0.33\pm0.03\pm0.03\pm0.03\pm0.02\pm0.01$&\  $0.075\pm0.002\pm0.001$
& \   $0.0039\pm0.0002\pm0.0001$ \\
$T_1^{B_c\to D^*_s}$  &\ $0.28\pm0.03\pm0.03\pm0.03\pm0.02\pm0.01$ &\  $0.064\pm0.001\pm0.001$
&\  $0.0035\pm0.0002\pm0.0001$  \\
$T_2^{B_c\to D^*_s}$  &\ $0.28\pm0.03\pm0.03\pm0.03\pm0.02\pm0.01$ &\  $0.039\pm0.001\pm0.001$
&\  $0.0023\pm0.0002\pm0.0001$  \\
$T_3^{B_c\to D^*_s}$  &\ $0.27\pm0.03\pm0.03\pm0.03\pm0.02\pm0.01$ &\  $0.082\pm0.002\pm0.001$
&\  $0.0068\pm0.0002\pm0.0002$  \\
\hline\hline
\end{tabular}
\end{center}
\end{table}

\begin{table}[thb]
\begin{center}
\caption{Comparison of $B_c\to D^{(*)}_{(s)}$ transition form factors
at $q^2=0$ evaluated in this paper with  other methods.}
\label{tab-FF-literature}
\begin{tabular}{l c c c c c c c c} \hline\hline
$B_c\to D^{(*)}$\; & $F_+(0)=F_0(0)$ &\; $F_T(0)$ \;
&\; $A_0(0)$ \; & \; $A_1(0)$ \; & \;  $A_2(0)$
&\; $V(0)$ \; & $T_1(0)=T_2(0)$ & \; $T_3(0)$  \\
pQCD & $0.19$ & $0.20$ & $0.17$ & $0.18$ & $0.20$
& $0.25$ & $0.22$ & $0.20$ \\
Ref.\cite{prd79-054012} & $0.16$ & $-$ & $0.09$ & $0.08$ & $0.07$
& $0.13$ & $-$ & $-$ \\
Ref.\cite{prd68-094020} & $0.14$ & $-$ & $0.14$ & $0.17$ & $0.19$
& $0.18$ & $-$ & $-$ \\
Ref.\cite{prd63-074010} & $0.189$ & $-$ & $0.284$ & $0.146$ & $0.158$
& $0.296$ & $-$ & $-$ \\
Ref.\cite{epjc51-833} & $0.35$ & $-$ & $0.05$ & $0.32$ & $0.57$
& $0.57$ & $-$ & $-$ \\
Ref.\cite{hepph-0211021} & $0.32$ & $-$ & $0.35$ & $0.43$ & $0.51$
& $1.66$ & $-$ & $-$ \\
Ref.\cite{prd79-304004} & $0.075$ & $-$ & $0.081$ & $0.095$ & $0.11$
& $0.16$ & $-$ & $-$ \\
\hline
$B_c\to D_s^{(*)}$\; & $F_+(0)=F_0(0)$ &\; $F_T(0)$ \;
&\; $A_0(0)$ \; & \; $A_1(0)$ \; & \;  $A_2(0)$
&\; $V(0)$ \; & $T_1(0)=T_2(0)$ & \; $T_3(0)$  \\
pQCD & $0.27$ & $0.28$ & $0.21$ & $0.23$ & $0.25$
& $0.33$ & $0.28$ & $0.27$ \\
Ref.\cite{prd79-054012} & $0.28$ & $-$ & $0.17$ & $0.14$ & $0.12$
& $0.23$ & $-$ & $-$ \\
Ref.\cite{hepph-0211021} & $0.45$ & $-$ & $0.47$ & $0.56$ & $0.65$
& $2.02$ & $-$ & $-$ \\
Ref.\cite{prd79-304004} & $0.15$ & $-$ & $0.16$ & $0.18$ & $0.20$
& $0.29$ & $-$ & $-$ \\
\hline\hline
\end{tabular}
\end{center}
\end{table}

With the form factors given, it is straightforward to calculate the branching ratios for all the
considered semileptonic decays by performing the numerical integration over the whole range of $q^2$.
For the $b\to u$ charged current process, with $l=(e, \mu)$, the decay rates are the following:
\beq\label{eq-br-cc}
Br(B^-_c\to \bar D^0l^-\bar\nu_l)&=&(3.15^{+0.97}_{-0.72}(\omega_{B_c})
^{+0.68}_{-0.61}(f_{B_c})^{+0.29}_{-0.27}(m_c)^{+0.31}_{-0.29}(C_D)
^{+0.28}_{-0.27}(f_D)\pm0.28(\tau_{B_c}))\cdot 10^{-5}, \non
Br(B^-_c\to \bar D^0\tau^-\bar\nu_\tau)&=&(2.16^{+0.72}_{-0.52}(\omega_{B_c})
^{+0.46}_{-0.42}(f_{B_c})^{+0.22}_{-0.19}(m_c)^{+0.20}_{-0.19}(C_D)
^{+0.19}_{-0.18}(f_D)\pm0.19(\tau_{B_c}))\cdot 10^{-5}, \non
Br(B^-_c\to \bar D^{*0}l^-\bar\nu_l)&=&(1.09^{+0.34}_{-0.26}(\omega_{B_c})
^{+0.23}_{-0.21}(f_{B_c})^{+0.13}_{-0.11}(m_c)\pm0.10(C_{D^*})
^{+0.23}_{-0.21}(f_{D^*})\pm0.10(\tau_{B_c}))\cdot 10^{-4}, \non
Br(B^-_c\to \bar D^{*0}\tau^-\bar\nu_\tau)&=&(0.64^{+0.20}_{-0.16}(\omega_{B_c})
^{+0.14}_{-0.12}(f_{B_c})^{+0.08}_{-0.07}(m_c)^{+0.06}_{-0.05}(C_{D^*})
^{+0.13}_{-0.12}(f_{D^*})\pm0.06(\tau_{B_c}))\cdot 10^{-4},
\eeq
where the errors come from the uncertainties of $\omega_{B_c}=1.0\pm0.1$,
$f_{B_c}=0.489\pm0.050$ GeV, $m_c=(1.275\pm0.025)$ GeV,
$C_{D^{(*)}}=0.5\pm0.1$, $f_D=(206.7\pm8.9)$ MeV or $f_{D^*}=(270\pm27)$ MeV
and $\tau_{B_c}=(0.45\pm0.04)$ ps, respectively.

For the flavor-changing neutral-current processes, after making the numerical
integration over the whole range of $4m_l^2\leq q^2\leq(m_{B_c}-m)^2$,
we get the pQCD predictions for the branching ratios
of considered decay modes which are listed in Table \ref{tab-Br-FCNC}.
The errors of the pQCD predictions in Table \ref{tab-Br-FCNC} come from
the uncertainties of $\omega_{B_c}$, $m_c$, $C_{D^{(*)}}$ or $C_{D^{(*)}_s}$,
$f_{D^{(*)}}$ or $f_{D_s^{(*)}}$ and $\tau_{B_c}$, respectively.

\begin{table}[thb]
\begin{center}
\caption{The pQCD predictions for the branching ratios of the considered
decays ($l=e,\mu$).}
\label{tab-Br-FCNC}
\begin{tabular}{l l } \hline\hline
\ Decay modes & pQCD predictions   \\
 \hline
 $Br(B^-_c\to D^-l^+l^-)$ \
&$(3.79^{+1.16}_{-0.86}(\omega_{B_c})^{+0.81}_{-0.74}(f_{B_c})^{+0.35}_{-0.32}(m_c)^{+0.37}_{-0.35}(C_D)
^{+0.33}_{-0.32}(f_D)\pm0.34(\tau_{B_c}))\cdot 10^{-9}$ \\
 $Br(B^-_c\to D^-\tau^+\tau^-)$ \
&$(1.03^{+0.38}_{-0.27}(\omega_{B_c})^{+0.22}_{-0.20}(f_{B_c})^{+0.12}_{-0.10}(m_c)^{+0.09}_{-0.08}(C_D)
\pm0.09(f_D)\pm0.09(\tau_{B_c}))\cdot 10^{-9} $\\
 $Br(B^-_c\to D^-\bar\nu\nu)$ \
&$(3.13^{+0.96}_{-0.71}(\omega_{B_c})^{+0.67}_{-0.61}(f_{B_c})^{+0.30}_{-0.26}(m_c)^{+0.31}_{-0.29}(C_D)
^{+0.28}_{-0.26}(f_D)\pm0.28(\tau_{B_c}))\cdot 10^{-8} $\\
 $Br(B^-_c\to D_s^-l^+l^-)$ \
&$(1.56^{+0.46}_{-0.36}(\omega_{B_c})^{+0.33}_{-0.30}(f_{B_c})^{+0.17}_{-0.15}(m_c)^{+0.13}_{-0.12}(C_{D_s})
\pm0.07(f_{D_s})\pm0.14(\tau_{B_c}))\cdot 10^{-7}$\\
 $Br(B^-_c\to D_s^-\tau^+\tau^-)$ \
&$(0.38^{+0.13}_{-0.10}(\omega_{B_c})^{+0.08}_{-0.07}(f_{B_c})^{+0.05}_{-0.04}(m_c)\pm0.03(C_{D_s})
\pm0.02(f_{D_s})\pm0.03(\tau_{B_c}))\cdot 10^{-7}  $\\
 $Br(B^-_c\to D_s^-\bar\nu\nu)$ \
&$(1.29^{+0.39}_{-0.30}(\omega_{B_c})^{+0.28}_{-0.25}(f_{B_c})^{+0.14}_{-0.12}(m_c)^{+0.11}_{-0.10}(C_{D_s})
\pm0.06(f_{D_s})\pm0.11(\tau_{B_c}))\cdot 10^{-6} $ \\
\hline
 $Br(B^-_c\to D^{*-}l^+l^-)$ \
&$(1.21^{+0.36}_{-0.28}(\omega_{B_c})^{+0.26}_{-0.23}(f_{B_c})^{+0.14}_{-0.12}(m_c)\pm0.11(C_{D^*})
^{+0.25}_{-0.23}(f_{D^*})\pm0.11(\tau_{B_c}))\cdot 10^{-8}$\\
 $Br(B^-_c\to D^{*-}\tau^+\tau^-)$ \
&$(0.16^{+0.05}_{-0.04}(\omega_{B_c})\pm0.03(f_{B_c})\pm0.02(m_c)\pm0.01(C_{D^*})
^\pm0.03(f_{D^*})\pm0.01(\tau_{B_c}))\cdot 10^{-8} $ \\
 $Br(B^-_c\to D^{*-}\bar\nu\nu)$ \
&$(1.10^{+0.34}_{-0.26}(\omega_{B_c})^{+0.24}_{-0.21}(f_{B_c})^{+0.13}_{-0.11}(m_c)\pm0.10(C_{D^*})
^{+0.23}_{-0.21}(f_{D^*})\pm0.10(\tau_{B_c}))\cdot 10^{-7} $\\
 $Br(B^-_c\to D_s^{*-}l^+l^-)$ \
&$(4.40^{+1.40}_{-1.05}(\omega_{B_c})^{+0.95}_{-0.85}(f_{B_c})^{+0.72}_{-0.57}(m_c)^{+0.32}_{-0.31}(C_{D^*_s})
^{+0.92}_{-0.84}(f_{D^*_s})\pm0.39(\tau_{B_c}))\cdot 10^{-7}$\\
 $Br(B^-_c\to D_s^{*-}\tau^+\tau^-)$ \
&$(0.52^{+0.18}_{-0.13}(\omega_{B_c})^{+0.11}_{-0.10}(f_{B_c})^{+0.10}_{-0.08}(m_c)\pm0.03(C_{D^*_s})
^{+0.11}_{-0.10}(f_{D^*_s})\pm0.05(\tau_{B_c}))\cdot 10^{-7} $\\
 $Br(B^-_c\to D_s^{*-}\bar\nu\nu)$ \
&$(4.04^{+1.30}_{-0.97}(\omega_{B_c})^{+0.87}_{-0.78}(f_{B_c})^{+0.68}_{-0.53}(m_c)^{+0.29}_{-0.28}(C_{D^*_s})
^{+0.85}_{-0.77}(f_{D_s^*})
\pm0.36(\tau_{B_c}))\cdot 10^{-6} $\\
\hline\hline
\end{tabular}
\end{center}
\end{table}

From the pQCD predictions for the form factors $F_{0,+,T}$ in Table \ref{tab-FF-I},
the form factors $V, A_{0,1,2}$ and $T_{1,2,3}$ in Table \ref{tab-FF-II} and the
pQCD predictions for the branching ratios as listed in Eq.~(\ref{eq-br-cc}) and in
Table \ref{tab-Br-FCNC}, we have the following points:
\begin{enumerate}
\item[(i)]
 All the form factors for the transitions $B_c\to D^{(*)}_s$ are  larger than
  the corresponding values for the transitions $B_c\to D^{(*)}$ at $q^2=0$, which characterizes the SU(3) breaking effect.

\item[(ii)]
  $F_0(0)$ equals to $F_+(0)$ by definition for the
  $B_c\to D$ or $B_c\to D_s$ transition, but they have different $q^2$ dependence
   by the different parameters $(a, b)$. $T_1(0)$ equals to $T_2(0)$ for the
  $B_c\to D^*$ or $B_c\to D^*_s$ transition claimed by the Eq. (\ref{eq-condition-t1t2})
  as they are given in Table \ref{tab-FF-II} although their expressions are
  different in Eqs.(\ref{eq:T1qq}, \ref{eq:T2qq}).

\item[(iii)]
  Because of the phase space suppression, the branching ratios of the decay modes with a
  $\tau$ in the final product are smaller than those decay modes with
  electron or muon in the final product for the the charged current
  process. And for the flavor changing neutral current processes, with two $\tau$'s in the
  final product, the branching ratios are much smaller than the corresponding decays
  with electron or muon pairs in the final product.

\item[(iv)]
  The branching ratios of the decay modes with $\bar\nu\nu$ are almost an order magnitude larger
  than the corresponding decays with $l^+l^-$ after the summation over
  the three neutrino generations. Because of the strong suppression of the CKM factor
  $|V_{td}/V_{ts}|^2=|0.211|^2$ \cite{pdg2012}, the branching ratios for the decay modes with
  $b\to d$ transitions are much smaller than those decay modes with the $b\to s$ transitions.
\end{enumerate}

  In order to reduce the theoretical uncertainty of the form factor calculations, we define
  two ratios $R_D$ and $R_{D^*}$ among the branching ratios for the the charged-current
  processes
  \beq
  R_D&=&\frac{Br(B^-_c\to \bar D^0\tau^-\bar\nu_\tau)}{Br(B^-_c\to \bar D^0l^-\nu_l)}
  = 0.69\pm0.01(\omega_{B_c})^{+0.01}_{-0.00}(m_c)\;,\\
  R_{D^*}&=&\frac{Br(B^-_c\to \bar D^{*0}\tau^-\bar\nu_\tau)}{Br(B^-_c\to \bar D^{*0}
  l^-\nu_l)}=0.59^{+0.00}_{-0.01}(\omega_{B_c})^{+0.00}_{-0.01}(m_c)\;,
  \eeq
  with $l=(e,\mu)$. These two relations will be tested by experiments.

\section{Summary } \label{sec-sum}

In this paper we studed  the $B_c \to (D_{(s)},D^*_{(s)})$ transition form factors $F_{0,+,T}(q^2)$
and $V(q^2), A_{0,1,2}(q^2), T_{1,2,3}(q^2)$ in the pQCD factorization approach based on $k_T$
factorization. The pQCD predictions for the values of the $B_c \to D_{(s)}$ and $B_c \to D^*_{(s)}$
transition form factors agree with those obtained using other methods. Utilizing these form factors,
we calculated the branching ratios for all the semileptonic decays of
$B_c^+\to D^{(*)}_{(s)}(l^+\nu_l,l^+l^-,\nu\bar\nu)$.
Because of phase space suppression, the production ratios of the decay modes with lepton $\tau$ in
the final product are smaller than the corresponding decays with electron or muon in the final product.
The branching ratios of the decay modes with $\bar\nu\nu$ are almost an order magnitude larger than
the corresponding decays with $l^+l^-$ after the summation over the three neutrino generations.
The branching ratios for the decays with the $b\to d$ transition are much smaller than those
with the $b\to s$ transitions.

In order to reduce the theoretical uncertainty of the pQCD predictions, we defined
two ratios $R_D$ and $R_{D^*}$ among the branching ratios for the the charged-current
processes. The pQCD predictions are
\beq
R_D&=&\frac{Br(B^-_c\to \bar D^0\tau^-\bar\nu_\tau)}{Br(B^-_c\to \bar D^0l^-\nu_l)}
\approx 0.7\;,\\
R_{D^*}&=&\frac{Br(B^-_c\to \bar D^{*0}\tau^-\bar\nu_\tau)}{Br(B^-_c\to \bar D^{*0}l^-\nu_l)}
\approx 0.6\;,
\eeq
with $l=(e,\mu)$. It would possible to test these predictions by LHCb and the forthcoming
Super-B experiments.

\begin{acknowledgments}
This work is supported in part by the National Natural Science
Foundation of China under Grant No. 11375208, 11228512, and 11235005.
\end{acknowledgments}



\end{document}